\documentclass[twocolumn,aps,showpacs,showkeys,prd,superscriptaddress,byrevtex]{revtex4-1}

\usepackage{todonotes}
\usepackage{bm}
\usepackage{graphicx}
\usepackage{amsmath}
\usepackage{color}
\usepackage{gensymb}

\makeatletter
\let\cat@comma@active\@empty
\makeatother

\begin{document}

\title{Correlating the nanoscale structural, magnetic and magneto-transport properties in SrRuO$_3$-based perovskite oxide ultra-thin films}

\author{Gerald Malsch}
\email{gerald.malsch@tu-dresden.de} 
\author{Dmytro Ivaneyko}
\email{dmytro.ivaneiko@tu-dresden.de}
\author{Peter Milde}
\email{peter.milde@tu-dresden.de}
\affiliation{TU Dresden, Institute of Applied Physics, 01062  Dresden, Germany}
\author{Lena Wysocki}
\author{Lin Yang}
\author{Paul H.M. van Loosdrecht}
\author{Ionela Lindfors-Vrejoiu}
\email{vrejoiu@ph2.uni-koeln.de}
\affiliation{II. Physikalisches Institut, Universit\"at zu K\"oln, 50937 K\"oln, Germany}
\author{Lukas M. Eng}
\email{lukas.eng@tu-dresden.de}
\affiliation{TU Dresden, Institute of Applied Physics, 01062  Dresden, Germany}
\affiliation{ct.qmat, Dresden-W\"urzburg Cluster of Excellence - EXC 2147, TU Dresden, 01062 Dresden, Germany}

\begin{abstract}
We investigated the structural and magnetic properties of bare SrRuO$_3$ (SRO) ultra-thin films and SrRuO$_3$/SrIrO$_3$/SrZrO$_3$ (SRO/SIO/SZO: RIZ) trilayer heterostructures between 10~K and 80~K, by comparing macroscopic data using magneto-optical Kerr effect (MOKE) and magneto-transport (anomalous Hall effect: AHE), with nanoscale fingerprints when applying non-contact scanning force microscopy (nc-SFM) and magnetic force microscopy (MFM). SRO and RIZ ultra-thin films were epitaxially grown at $650^\circ\text{C}$ onto vicinal SrTiO$_3$ (100) single-crystalline substrates to a nominal thickness of 4 and 4/2/2 unit cells (uc), respectively. Our correlated analysis allows associating topographic sample features of overgrown individual layers to their residual magnetization, as is shown here to be relevant for interpreting the macroscopic AHE data. Although the hump-like features in the AHE suggest a magnetically textured skyrmion phase to exist around 55~K associated to the topological Hall effect (THE), both our MOKE and MFM data cannot support this theory. In contrast, our SFM/MFM local-scale analysis finds the local coercive field to be strongly dependent on the effective layer thickness and stoichiometry in both the SRO and RIZ samples, with huge impact on the local band-structure. In fact, it is these variations that in turn mimic a potential THE through anomalies in the AHE resistivity loops.
\\
\end{abstract}

\maketitle 

\section{Introduction}
Magneto-transport and Hall measurements are versatile techniques to inspect and characterize magnetically active materials on the micrometer length scale. Extra contributions to the ordinary Hall effect arising through the sample magnetization dependence, have been identified as the {\it so-called} anomalous Hall effect (AHE). In ferromagnetic materials, the AHE usually is proportional to the sample magnetization and therefore shows the same hysteretic behavior~\cite{Karplus54,Nagaosa10}.

Recently, additional contributions to the AHE have been postulated that arise from topologically non-trivial magnetic textures~\cite{Bruno04}. This component was labelled as the {\it Topological Hall Effect (THE)} and is of great importance when proving the existence of skyrmions by simple transport measurements. As a non-trivial topological magnetic texture, skyrmion lattices (SkLs) are expected to contribute to the THE, being visible as hump-like anomalies on top of the expected AHE~\cite{Neubauer09}. As a conclusion, it is often assumed that any contribution resembling the THE recognized in the Hall data might originate from the presence of skyrmions~\cite{Matsuno16}.

The research field in skyrmion phenomena is given a lot of attention these days, specifically also with the big view of finding the next-generation material that supports skyrmion formation, especially for data storage applications~\cite{Fert13}. One avenue of research is to discover an insulating skyrmion host material, that in turn, would allow electric field manipulation of magnetic domains/phases using standard gating techniques~\cite{Mizuno17,Ohuchi18}. Hence, epitaxially-grown ferromagnetic perovskite oxide heterostructures are of central interests~\cite{Matsuno16,Ohuchi18,Sohn18,Wang18,Meng19} since they undergo a metal-to-insulator transition for layer thicknesses measuring between 3 and 6 unit cells~\cite{Kan18}.

A ferromagnetic perovskite oxide that is well suited for generating skyrmions, is the 4d-transition metal SrRuO$_3$ (SRO). Epitaxial layers of SRO and heterostructures involving SRO-layers recently have been in the research focus, because of peculiar features appearing in the AHE resistivity loops that hint towards a significant THE contribution possibly inferred by skyrmions. This interpretation is currently under heated debate~\cite{Kan18,Wu19}. Therefore, we combine here magneto-transport with both nanometer- and micrometer-scale spectroscopy/microscopy for inspecting two selected samples, a bare SrRuO$_3$ ultra-thin film of a 4uc-thickness (4SRO) and a 4SRO layer overgrown with 2-by-2 uc of SrIrO$_3$ and SrZrO$_3$, resulting in the trilayer 4RIZ heterostructure. Our study unambiguously proves that the origin for the observed THE lies within the 4SRO layer, needing absolutely \textbf{no skyrmions} to be present in order to mimic the hump-like features observed in the THE data.

\section{Material}
SrRuO$_3$ (SRO) constitutes one of the few functional perovskite materials with formula ABO$_3$ that is ferromagnetic~\cite{Zhu14,Kundu16}. The ferromagnetic order is robust and preserved down to at least 3 unit cells (uc) when epitaxially grown on SrTiO$_3$ (STO) (100) single crystals~\cite{Matsuno16}. Furthermore, SRO films have a large magneto-crystalline anisotropy~\cite{Snyder19} that is easily manipulated for instance through epitaxial heterostructuring. At ordered perovskite interfaces, strong structural coupling via the oxygen octahedra provides a valuable way to engineer magnetic anisotropy, by controlling the easy-axes orientation of the magnetization~\cite{Kan15,Kan16}. Magnetic interfacial coupling also provides an appropriate route to manipulate magnetic ordering, for instance in SrRuO$_3$/La$_{0.7}$Sr$_{0.3}$MnO$_3$ superlattices engineered for strong anti-ferromagnetic interlayer coupling that results in non-collinear magnetic ordering~\cite{Ziese12,Kim12}. 

In 1999, a Lorentz transmission electron microscopy study~\cite{Marshall99} reported on 30-100-nm-thick SRO films revealing magnetic stripe domains that are separated by 3-nm-narrow domain walls (DWs). These domains drastically impact the linear magnetoresistance for zero-field cooled SRO films, as documented by Klein {\it et al.}~\cite{Klein98}. A more recent study reports on a 4-K MFM investigation of patterned rectangular magnetic nano-islands of a 10-nm-thick SRO film~\cite{Landau12}.

With respect to skyrmion formation in heterostructures based on broken inversion symmetry, SRO is a very attractive candidate due to the perpendicular magnetic anisotropy, exhibited when epitaxially grown on single crystalline (sc) STO(100). A very recent example is the work on SrRuO$_3$/SrIrO$_3$ (SRO/SIO) bilayers on STO(100) substrates, in which the ferromagnetic ultra-thin SRO layer is interfaced with the paramagnetic SIO that exhibits large spin-orbit coupling (SOC)~\cite{Matsuno16}. The inherent inversion symmetry breaking at the interface as well as the proximity of the large SOC of heavy 5d Ir ions, are expected to result in a strong interfacial Dzyaloshinskii-Moriya interaction (DMI). The latter may lead to non-collinear magnetic ordering in the ultra-thin SRO layers (around 4 - 6 unit cells) with strong perpendicular magnetic anisotropy~\cite{Matsuno16,Ohuchi18}. 

Matsuno {\it et al.} showed MFM investigations of a 5uc-SRO/2uc-SIO bilayer heterostructure that were claimed to hint towards the formation of tiny bubble-like magnetic domains, interpreted as a possible skyrmion phase~\cite{Matsuno16}. Therefore, we investigate here the magnetic domain formation and morphology in epitaxial ultra-thin SRO films and SRO/SIO/SZO (RIZ) heterostructures, in which the SRO layer thickness is kept the same (i.e. 4uc). We find the nanoscopic origins of SRO layer overgrowth as well as the locally varying coercive field within the 4uc SRO films to be the origin of macroscopic AHE anomalies, rather than a textured skyrmion phase.

\section{Sample preparation}
\begin{figure}[!h]
\includegraphics*[width=\columnwidth]{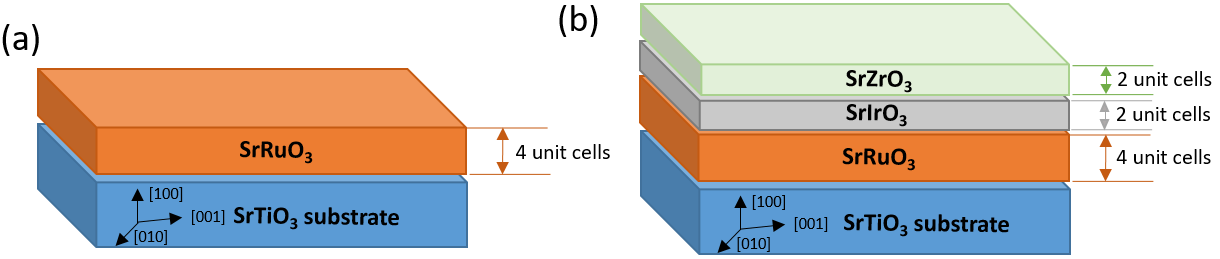}
\caption{\textbf{Sample design} of (a) the bare SRO thin film with a 4 pseudo-cubic unit cell thickness (4SRO), and (b) the 4RIZ SRO/SIO/SZO trilayer heterostructure made up from 4/2/2 unit cells, respectively.}
\label{fig001}
\end{figure}

The heterostructures under study [see Fig.~\ref{fig001}] were fabricated by pulsed-laser deposition (PLD) using a KrF excimer laser. High-oxygen pressure reflective high energy electron diffraction (RHEED) was used for both monitoring the layer-growth {\it in-situ} and to analyze the film structure at the growth temperature and after cooling (see also supplemental information S1). sc-STO(100) vicinally cut under an angle of about $0.1^\circ\text{C}$ were used as substrates after etching in buffered HF and annealing at $1000^\circ\text{C}$ for 2 hours in air. These substrates then show a uniform and continuous TiO$_2$ surface termination, hence providing a stepped sample surface with regularly distributed terraces of a 200 - 450~nm width.

Stoichiometric SRO, SIO, and SrZrO$_3$ (SZO) targets were employed for PLD. The layer growth was performed in a 0.133~mbar O$_2$ atmosphere, and STO substrates were heated to $650^\circ\text{C}$. The laser fluence was set to 2~J/cm$^2$ while the laser repetition rate for the SRO and SIO/SZO were 5 and 1-2~Hz, respectively. SZO was used here as a protective capping layer on top of the SRO/SIO bilayer, in order to avoid deterioration in the SIO stoichiometry due to moisture~\cite{Groenendijk16}. Both the SIO and SZO were grown to a 2 pseudo-cubic uc thickness each, as determined by RHEED specular spot oscillations. As a result, we obtained a set of 2 samples [see Fig.~\ref{fig001}]: (a) a 4uc-thin bare SRO sample (furtheron labelled as {\it 4SRO}), and (b) a 4/2/2-uc thin SRO/SIO/SZO trilayer heterostructure (labelled furtheron as the {\it 4RIZ} sample).

\noindent

\section{Magnetic and magneto-transport studies}
4SRO and 4RIZ samples firstly were analyzed at the macroscopic length scale, employing the magneto-optical Kerr effect (MOKE) (for details see supplement S2), and magneto-transport measurements using a van-der-Pauw Hall setup in transverse geometry. In order to suppress the contribution of the linear magnetoresistance, electrical contacts were cyclically permutated to read proper Hall data. 

It is commonly assumed that the total Hall resistivity of a ferromagnet is given by the sum of the ordinary Hall resistivity $\rho_{OHE}$ and the anomalous Hall resistivity contribution $\rho_{AHE}$ as:
\begin{equation}
\rho_{xy}=\rho_{OHE} +  \rho_{AHE} = \mu_0 (R_0 H_z + R_A M_z).
\label{eq01}
\end{equation}

\begin{figure}[!b]
\includegraphics*[width=\columnwidth]{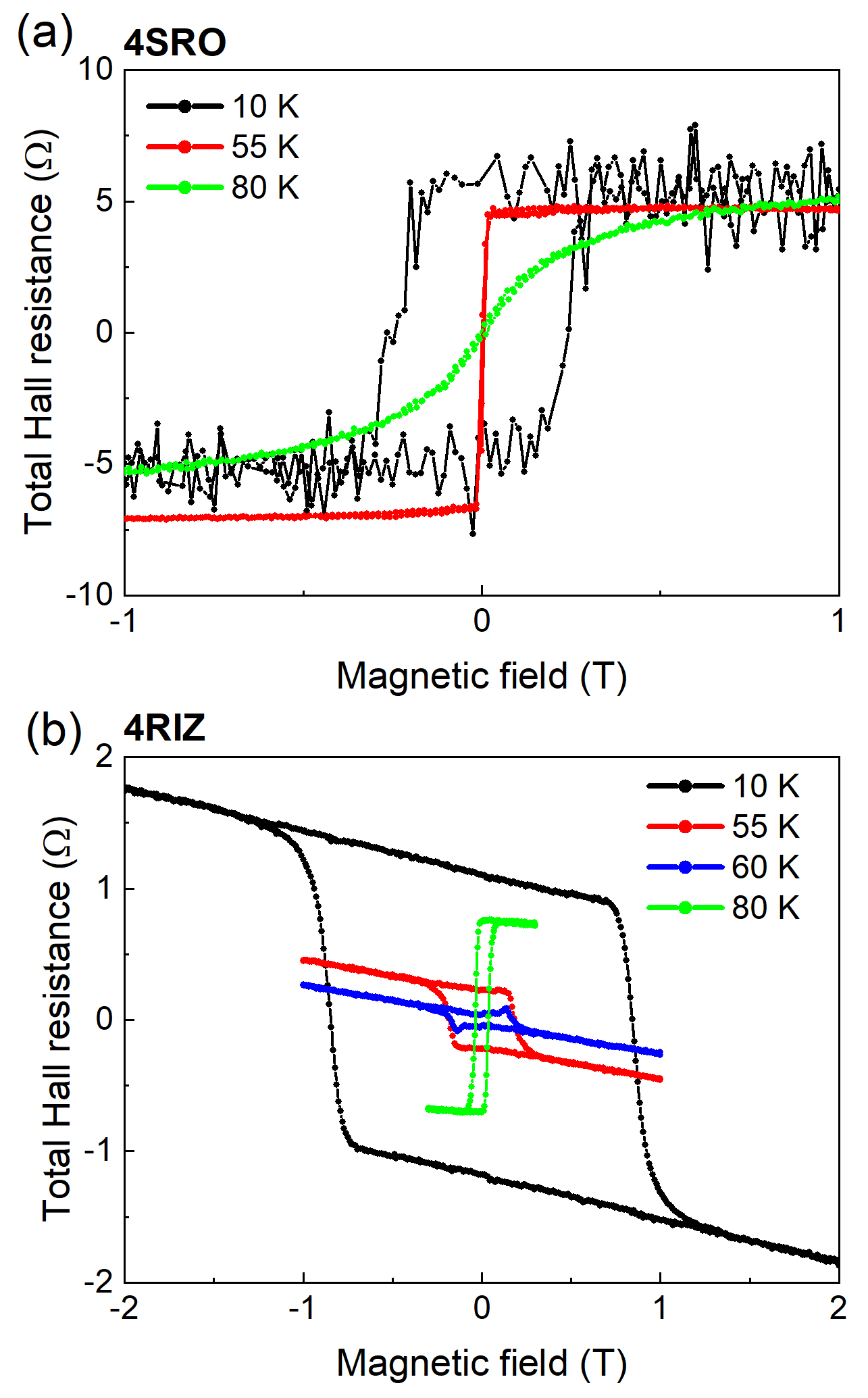}
\caption{\textbf{Total Hall resistance loops} for (a) the 4SRO film, and (b) the 4RIZ trilayer heterostructure. Note that the 4SRO in (a) is paramagnetic above 80 K, while the 4RIZ heterostructure shows an open $\rho_{xy}$ hysteresis loop proving the ferromagnetic order.}
\label{fig_Total_Hall_comparison}
\end{figure} 
\noindent 

Here $\mu_0$ is the permeability of vacuum, $R_0$ is the Hall coefficient that mainly  depends on the majority-charge carrier density, and $H_z$ denotes the magnitude of the external magnetic field applied perpendicular to the $xy$- sample plane. The second term in Eq.~(\ref{eq01}) accounts for the anomalous Hall effect exhibited in the presence of a spontaneous magnetization and large spin-orbit coupling; the AHE resistivity is directly proportional to the macroscopic magnetization component ${M_z}$ perpendicular to the current flow direction~\cite{Pugh1953}, with $R_A$ denoting the anomalous Hall coefficient~\cite{Nagaosa10}.

Fig.~\ref{fig_Total_Hall_comparison} displays the magnetic field dependence of the total Hall resistance as measured for (a) the bare 4SRO thin film, and (b) the 4RIZ trilayer heterostructure at selected temperatures. Capping the nominally 4SRO layer with 2-by-2 layers of the strong SOC SIO and the large bandgap insulator SZO, results in dramatic changes in both the magnetic and magneto-transport behavior of the SRO layers.

Bare 4SRO thin film exhibits a positive anomalous Hall effect constant $R_A$ within its ferromagnetic phase [Fig.~\ref{fig_Total_Hall_comparison}(a)], while the 4RIZ trilayer shows a negative $R_A$ at low temperatures switching to positive values close to 70~K. [Fig.~\ref{fig_Total_Hall_comparison}(b)]. 

Due to the previously found dependence of the anomalous Hall effect of SRO on the details of its band structure~\cite{Fang03}, the observed variations of the AHE characteristics may be related to differences in the crystal structure of the SRO layers~\cite{Ziese11}. The non-monotonous AHE temperature dependence already reported for a broad variety of SRO single crystals and epitaxial films~\cite{Ziese11,Fang03} including our 4RIZ trilayer structure, is a clear indication of structural stabilization of orthorhombic SRO layers by the overgrown layers of orthorhombic SIO and SZO. The stabilization of a tetragonal structure due to the suppression of the RuO$_6$ octahedra tilts was observed for SRO layers grown on DyScO$_3$(110) substrates when being capped with the cubic SrTiO$_3$ with no tilts of the oxygen octahedra, with direct impact on the Curie temperature of the ferromagnetic SRO film~\cite{Thomas17}. In case of our trilayer sample, the ferromagnetic transition temperature of the underlying SRO is significantly enhanced.

The Hall resistance loops of the 4SRO films at 80~K exhibits a $S$-shaped behavior, indicating that the layer must already be in its paramagnetic state. However, at 80~K, the Hall loop of the 4RIZ trilayer still displays an open hysteresis curve, which demonstrates that the SRO layer here is ferromagnetic, in agreement with our MFM investigations (see later).

In strong magnetic fields where the sample magnetization is assumed to be constant, the total Hall resistance of the 4RIZ trilayer reveals a negative slope that is attributed to electron-dominated Hall transport [see Fig.~\ref{fig_Total_Hall_comparison}(b)]; moreover, the corresponding fluctuations in the Hall resistivity of the 4SRO layer [Fig.~\ref{fig_Total_Hall_comparison}(a)] are negligibly small. The observed differences of the transport properties hence clearly indicate severe changes in the electronic band structure for the SRO layer of the 4RIZ trilayer structure.

\begin{figure}[!tb]
\includegraphics*[width=\columnwidth]{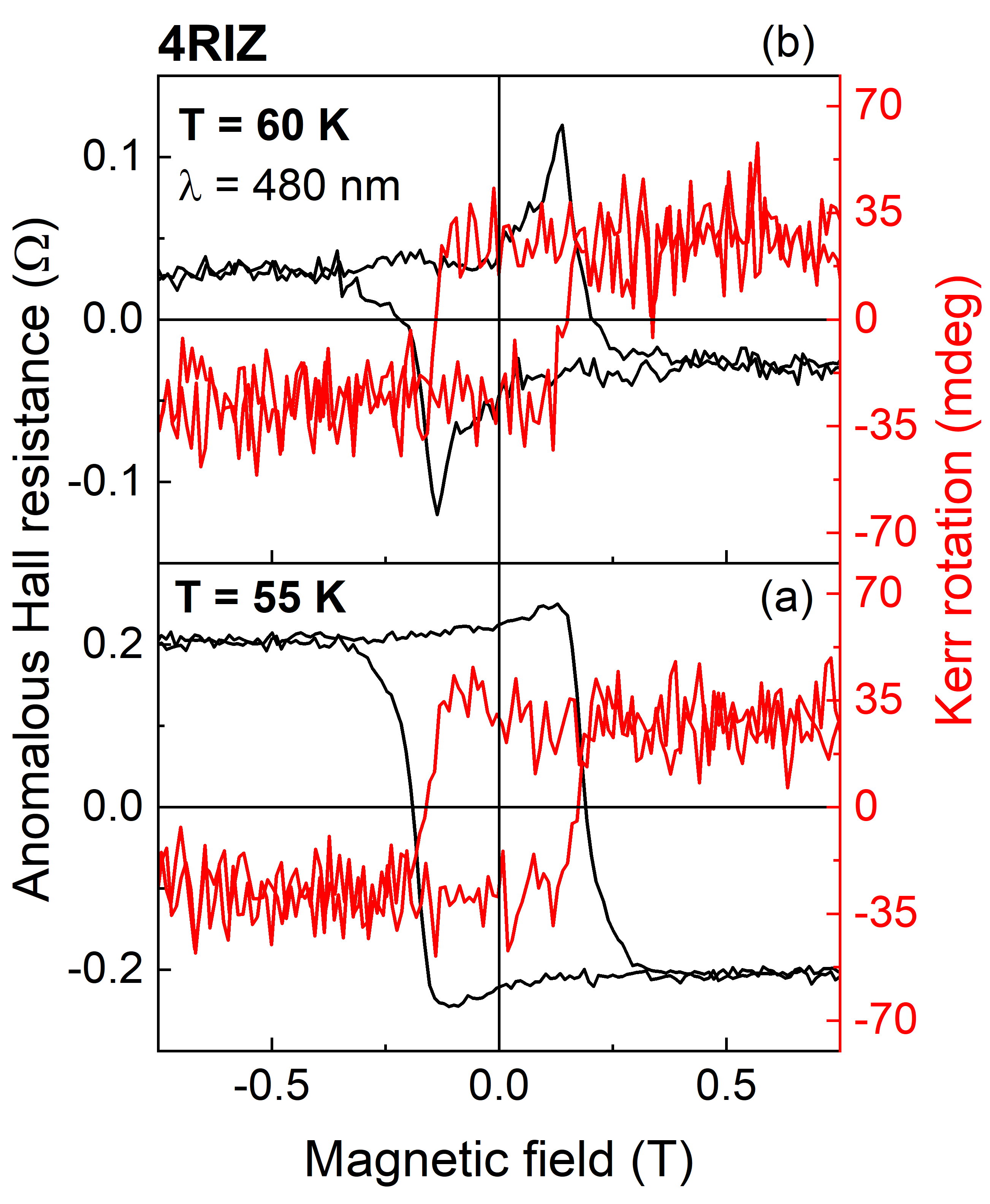}
\caption{Magnetic field dependence of the\textbf{ anomalous Hall resistance} (black) and the \textbf{Kerr rotation angle} (red) for the 4RIZ trilayer sample measured close to the AHE compensation temperature at 55 K (a) and 60 K (b), respectively. MOKE measurements were carried out with incoherent light at ${\lambda}$ = 480 nm in order to suppress optical artefacts in these measurements.}
\label{fig_Kerr_Hall}
\end{figure} 

Close to the temperature where the AHE changes sign (around 55~K), hump-like anomalies pop up in the Hall resistance when investigating our 4RIZ trilayer [see Fig.~\ref{fig_Total_Hall_comparison}(b)], in fact mimicking a contribution that might be attributed to the THE. Recently, several works claim the observation of a THE contribution to the AHE resistivity hysteresis loops in SRO ultrathin films and SRO/SIO bilayers~\cite{Matsuno16,Ohuchi18,Qin19,Wu19}. Notably, no such features were detected in our bare 4SRO layers. 

To clarify such an extra contribution, we compare our AHE with MOKE data obtained {\it in-situ} using a home-built MOKE setup (for details see supplementary information S2). Firstly, when sweeping the external magnetic field, the macroscopic MOKE rotation loops recorded on the 4RIZ thin film, resemble the switching behavior of a hard ferromagnetic layer in the easy axis configuration, independent of temperature [see Fig.~\ref{fig_Kerr_Hall}]. In fact, MOKE delivers no hints for the existence of an additional nontrivial magnetic phase, as inferred by the Hall measurements. Moreover, the coercive field strength and the magnetic field where the AHE vanishes, differ drastically [see Fig.~\ref{fig_Kerr_Hall}(b)]. 

These strong discrepancies hence can be solved only with a proper analysis of domain nucleation in SRO thin films, and shedding light on those nanoscale processes that unambiguously contribute to magnetization reversal in the 4RIZ heterostructures. We therefore apply low-temperature (LT) non-contact scanning-force microscopy (nc-SFM) and magnetic-force microscopy (MFM) to correlate the macroscopic AHE findings with local-scale structural and magnetic information on both the 4SRO and the 4RIZ samples, respectively. SFM/MFM was performed over the full temperature range from 10 to 80 K; the 55-K-results are discussed within the main text here, while all other data can be found in the supplement (see chapter S3).

\section{Low-temperature SFM and MFM measurements}
The nanoscale analysis of the two sample systems 4SRO and 4RIZ is carried out with our low-temperature (LT) non-contact scanning force microscope (SFM) operating under ultrahigh vacuum (base pressure below $ 2\cdot10^{-10} $ mbar). SSS-QMFMR-type MFM probes with a hard magnetic coating were used for both topographic (SFM) and magnetic (MFM) inspection, revealing a mechanical quality factor of $Q \geq  1.45\cdot10^5$. The nominal cantilever oscillation amplitude of 10~nm is kept constant at all times, while then taking the measured frequency shift $\Delta f$ for topographic and magnetic feedback control in SFM and MFM, respectively. 

MFM is performed in a two-path-mode, quantifying the sample topography in the first scan while then retracting the tip by 20~nm for the second scan, in order to be sensitive to the longer-ranged magnetic forces, only~\cite{Kazakova19}.


\subsection{Nanoscale real-space analysis of 4SRO and 4RIZ}
\begin{figure}[!tb]
	\includegraphics*[width=\columnwidth]{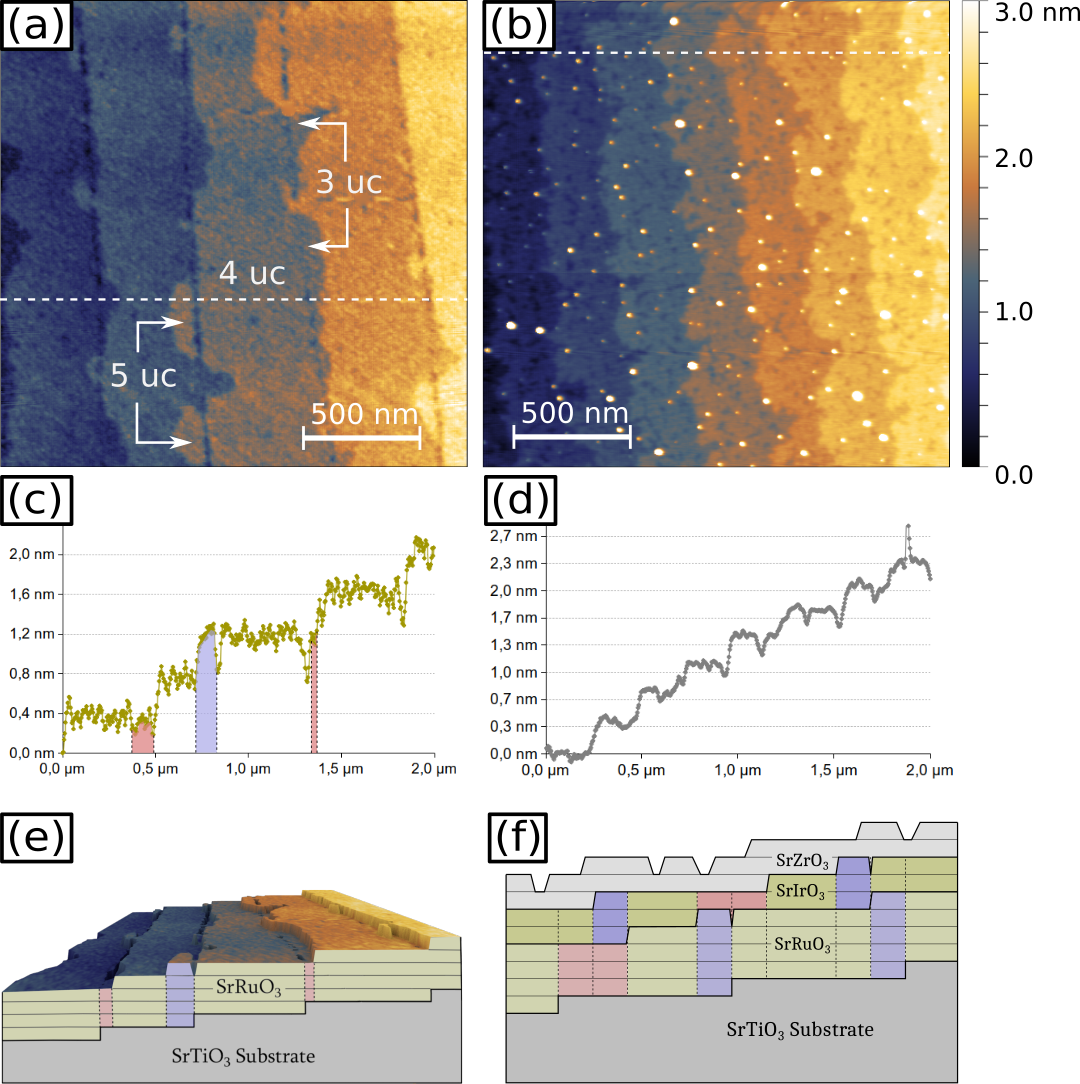}
	\caption{\label{figTopo} \textbf{Topographic images of (a) the bare 4SRO film, and (b) the 4RIZ trilayer heterostructure.} Note the over- and undergrown 5uc and 3uc areas in (a). Profiles of the stepped terraces in (c) 4SRO, and (d) the 4RIZ heterostructure, are taken along the dashed lines in (a) and (b), respectively. (e) Illustration of the differences in layer thickness as caused by under- and overgrowth of 4SRO. (f) Illustration of possible over- and undergrown combinations for the 4RIZ structure.}
\end{figure}

Fig.~\ref{figTopo} displays typical topographic nc-SFM scans of both the 4SRO (a) and the 4RIZ (b) samples. We clearly observe the stepped surface morphology with terraces of a 0.39~nm height extending over 200 - 450~nm each, as induced by the vicinal STO substrate. Due to the stochastic nature of the deposition process, step edges of the 4SRO surface do not exactly terminate at straight STO step edges, creating areas of sample over- and undergrowth by 5uc and 3uc of SRO, respectively. Fig.~\ref{figTopo}(a),(b) show this behavior for the two samples in a birds-view illustration ($xy$-scan), while Fig.~\ref{figTopo}(c),(d) and Fig.~\ref{figTopo}(e),(f) display the same fact in a cross-sectional and pseudo 3D manner.

\begin{figure*}[!bt]
	\includegraphics[width=\textwidth]{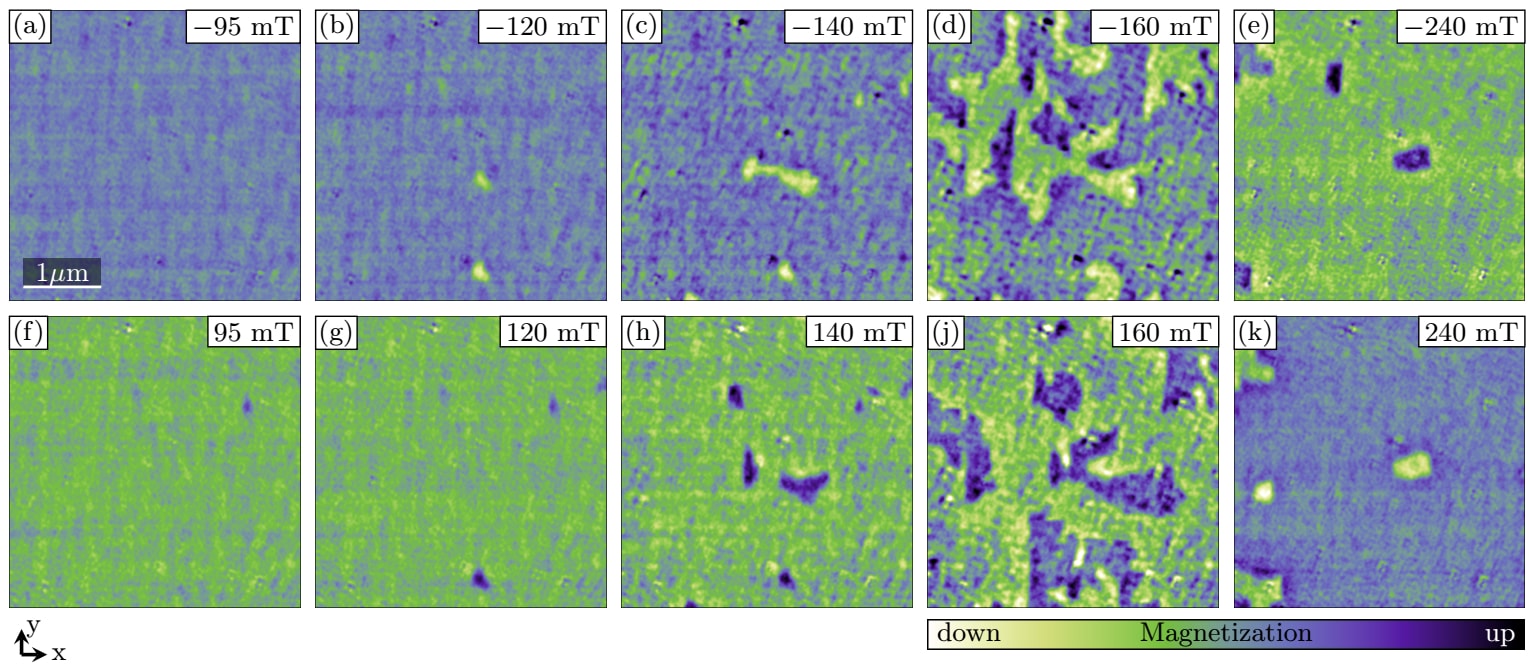}
	\caption{\label{figMagHyst55} \textbf{MFM measurements of the 4RIZ trilayer heterostructure at 55 K}, showing the domain formation during a forward [(a)-(e), upper row] and reversed magnetic field sweep [(f)-(k), lower row], respectively. Note that all magnetic features laterally extend over many terraces, reaching sizes of micrometers.}
\end{figure*}

The 4RIZ structure is presented in Fig.~\ref{figTopo}(b). This topographic scan shows a considerably more disturbed surface arrangement, with both several nm high peaks and many dips in the stepped surface, reaching depths comparable to the observed step height. Unlike for the bare 4SRO film, the discrimination of different layer heights is no longer possible, since SRO, SIO, and SZO layers intermix in numerous stacking variants, as illustrated in Fig.~\ref{figTopo}(f).

Step-and-terrace morphology that is typical to films grown in the PLD step-flow growth regime on vicinal substrates, turns out to play a crucial role in crystallographic domain formation and growth~\cite{Gan99,Ziese10}. Preferential alignment with terrace step edges oriented along the [001] orthorhombic axes has been reported for thinner SRO films, while rotation of the magnetic easy axis into the substrate plane is observed for medium-range film thicknesses (between 3 and 7.5~nm)~\cite{Schultz09}. Both our 4SRO and 4RIZ films clearly belong to the first class and scenario, concluding that the easy axis of the 4uc SRO layer always stands perpendicular to the sample surface. This is of great importance also to our MFM data acquisition and interpretation. 

\subsection{MFM of the 4RIZ trilayer heterostructure}
MFM/nc-SFM was applied for inspecting both the 4SRO and 4RIZ thin films at all temperatures. The 4SRO data at 10 K, 55 K, and 80 K are found in the supplemental chapters S3.1, S3.2, and S3.3, respectively, while the 4RIZ MFM inspections are posted in chapters S3.4 and S3.5 for 10 K and 80 K. The 55 K RIZ findings are discussed here and now.

A series of relevant MFM images when switching the 4RIZ trilayer heterostructure at 55~K from +2~T to -2~T are illustrated in Fig.~\ref{figMagHyst55}. The terraced substrate sample surface is still vaguely visible, aligned along the y-axes in all images. When sweeping from high magnetic fields at +2~T [Fig.~\ref{figMagHyst55}(k)] to the zero field state, the 4RIZ sample remains fully magnetized and no domains with reversed magnetization form. This is in full accordance to our hysteresis magnetization loops where the remanent magnetization is equal to the saturated magnetization, and consistent with literature~\cite{Xia09}. 

\begin{figure}[!b]
\includegraphics*[width=\columnwidth]{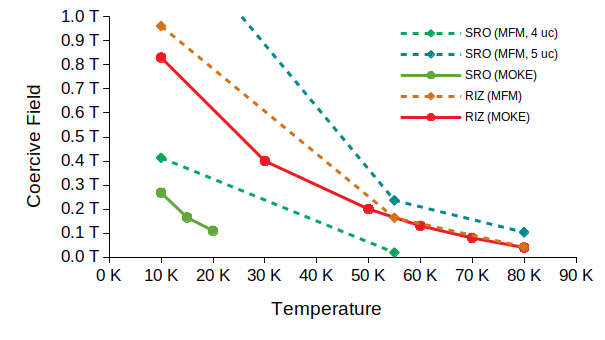}
\caption{\label{CoerciveFields} \textbf{ Coercive fields of 4SRO and 4RIZ samples}, and their areas depending on the number of unit cells. 
}
\end{figure}

Nevertheless, we still observe a low contrast visible along the step edges, due to differences in the magnetic susceptibility of different layer heights that in turn affects the MFM signal strength. At the negative field value of -22~mT, the tip changes its magnetization direction and aligns with the externally applied field.

Domain growth initiates at approximately -100~mT, forming small magnetic nuclei that rapidly increase their domain size when increasing the magnetic field; as shown, the film has mostly switched at -240~mT [see Fig.~\ref{figMagHyst55}(e)], with some isolated areas remaining pinned and stable up to -300~mT.

When reversing the field, we observe strong pinning of both initial magnetic nuclei and the switch-resistant areas, with several sites and domain wall shapes surviving over larger field changes and field ramping. One also sees that domain growth is affected, but not dominated, by the stepped substrate topography, with sharp vertical domain walls parallel to the terrace direction ($y$-axes) appearing during switching, a clear indication of realignment of the SRO to the topographically-induced anisotropy. Nonetheless, as grown/switched domains extend over multiple terraces reaching some micrometers in lateral size. 

By using Otsu's thresholding method~\cite{Otsu79} to separate the domains (for details see supplement S4), it is possible to extract hysteresis curves from individual MFM images, obtained for both the 4SRO and 4RIZ samples at different temperatures. The temperature dependence of the coercive field is shown in Fig.~\ref{CoerciveFields} for comparison with MOKE measurements. The width, slope and coercive fields of MFM hysteresis curves are in good agreement to the hysteresis curves observed by MOKE, and therefore also considerably sharper than the peaks observed in the Hall data. At the critical field where the hump-like features start to appear in the AHE, the 4RIZ sample still appears uniformly magnetized under MFM. At no point do we see an intermediate state or second transition indicating the presence of a magnetic structure different from the uniformly polarized domains, skyrmionic or otherwise.

\subsection{MFM of bare 4SRO thin films}
\begin{figure}[!b]
	\includegraphics*[width=\columnwidth]{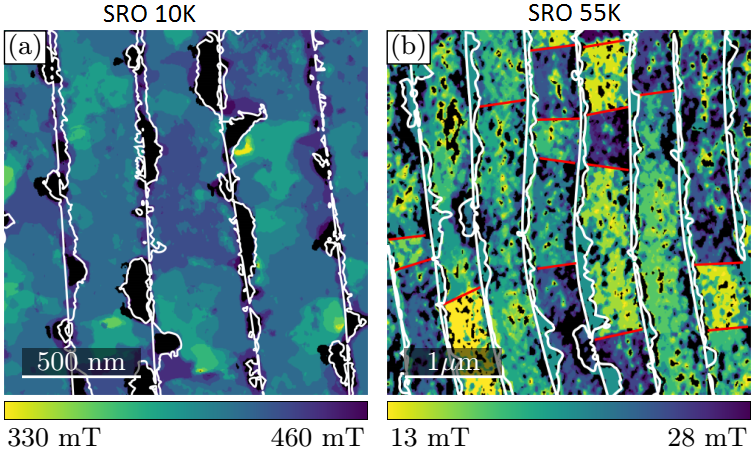}
	\caption{\label{CoerbyPix} \textbf{Local switching field of the 4SRO ultra-thin film}, measured at (a) 10~K and (b) at 55~K. Black areas do not switch within the shown interval. Note the large variation in $H_c$ over the sample areas at nominally the same thickness of 4uc. This variation is strongly temperature dependent.}
\end{figure}

For comparison reasons, field sweeps of the simpler 4SRO sample were as well investigated by MFM. While full MFM scans at all temperatures of 10~K, 55~K, and 80~K are found in the supplemental sections S3.1, S3.2, and S3.3, the summarized results for the 10~K and 55~K MFM findings on the 4RIZ thin films are illustrated in Fig.~\ref{CoerbyPix} here. 

What is plotted in Fig.~\ref{CoerbyPix} is a 2-dimensional map that shows the variation in the average magnetic field value needed to locally switch the sample surface at exactly that sample surface spot. We run a full hysteresis loop at every position of this 512 pixel x 512 pixel image, and then deduce (by applying Otsu's method) the corresponding local coercive field.

Notably, provided the thin-film sample is homogeneous, defect-free, and the growth being independent of any (topographic) structure, the resulting picture should reveal a uniform color/gray shade, since all areas/domains must switch at the same coercive field value, independent of temperatures. Concurrently, the error bar would be zero. Nevertheless, what we experimentally observe and display in Fig.~\ref{CoerbyPix}, is that even bare 4SRO is far from being ideal, in that the material switches unevenly: To the one side, the coercive field shows strong variations when plotted over the whole sample area, ranging from 13 to 28~mT at 55~K [Fig.~\ref{CoerbyPix}(b)] with an error bar of 15~mT, and 330 to 460~mT at 10~K [Fig.~\ref{CoerbyPix}(a)] with a 130~mT error bar. Rectangular areas along step edges are seen to switch cooperatively as a whole. Note also the black areas in these two figures that correspond to the 5uc and 3uc over-/undergrown areas. 

Also note from Fig.~\ref{CoerbyPix} that larger patches displayed in uniform color indicate areas that switch at the same coercive field, as nicely seen for very low temperatures at 10~K [Fig.~\ref{CoerbyPix}(a)]. All the greenish areas in Fig.~\ref{CoerbyPix} thus are 4uc SRO layers, and hence are expected to show the same switching behavior on the uniformly TiO$_2$-terminated STO (100) sample surface. Nevertheless, we find these areas to split up into SRO regions that have completely different switching fields, both parallel and perpendicular to the step edges. Variation of the epitaxial strain over the length scale of a terrace width is highly unlikely, which leaves us to conclude that electronic band structure variations for these different patches must be the origin for the documented variable switching behavior.

\begin{figure}[!b]
	\includegraphics*[width=\columnwidth]{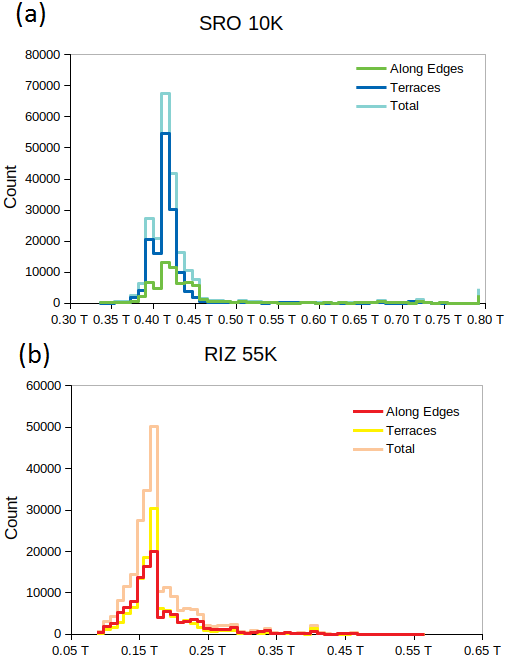}
	\caption{\label{Histograms} \textbf{Histograms of the coercive fields per pixel}, for the bare 4SRO thin film at 10~K (a) and the 4RIZ trilayer heterostructure at 55~K (b).}
\end{figure}

Also, the underlying anisotropy of the material causes the domains to elongate first perpendicular to the step edges before then growing along them. While at 10~K, this shape asymmetry becomes less pronounced and domain growth is observed both along and perpendicular to the STO step edges [see Fig.~\ref{CoerbyPix}(a)], the material also shows a strong preference for hosting nucleation sites, most preferentially at terrace step edges. This is an indicator that despite the large difference in coercive fields between the 4SRO and 4RIZ sample, the observed memory effect in the domain growth of the heterostructure is due to these defects within continuous 4SRO layers.

These field maps can be used for further analysis. We created histograms of areas adjacent to and far from the observed topographic step edges for the 4SRO-10-K and 4RIZ-55-K measurements to try and correlate changes in the field with these features. The results can be seen in Fig.~\ref{Histograms}: In both cases, the histograms close to the edges are slightly, but not significantly wider than on the terraces. For SRO, the distribution close to the edges shows a higher number of points switching later, while for RIZ, there is no obvious tendency to either direction. This is not an artefact of the larger scan range used for the RIZ measurements: If the SRO pictures are downsampled to feature a similar pixel resolution per $\mu$m, the resulting histograms retain their shapes and features.

In fact, one might have the impression that domains monitored at an early stage of nucleation or at the later annihilation stage, do resemble isolated skyrmion bubbles such as the ones reported for metallic multilayer systems Pt/Ir/Co~\cite{Moreau-Luchaire16}. However, our MFM investigations clearly cannot support this picture: 

\begin{itemize}
\item Firstly smallest feature sizes that could possibly be interpreted as {\it skyrmions} are of a $\sim10$~nm size only, but mostly at the resolution limit of our MFM investigations. 
\item Secondly, these bubble-like features occur quite rarely, and always as isolated structures that never form a continuous SkL.
\end{itemize}

Our MFM investigations hence cannot support the presence of any skrymions in these samples, neither isolated skyrmions nor any SkL, as we had found in many other single crystalline materials~\cite{Milde13,Kezsmarki15,Milde16,Zhang16,Butykai17,Neuber18}. What seems much more likely though, is that the local sample inhomogeneities (stoichiometry, band-structure) reflected through the varying local coercive fields, are the origin for mimicking the THE behavior in our transport data. The anomalous Hall constant $R_A$ in turn varies slightly and impacts the AHE loops at 55 K and 60 K to differ from both the Kerr loops and local-scale magnetization loops acquired by MFM.

\section{Conclusions}
Our correlated study involving MOKE, nc-SFM, MFM, and Hall transport proves that subtle differences in both the magnetization and structural overgrowth of the 4uc SRO layered thin film are the origin for both macroscopic transport anomalies and nanoscopic magnetic hysteresis fluctuations. MOKE and MFM results are extremely consistent when analyzing both type of samples, the bare 4uc SRO film PLD grown on vicinal STO(100), and the 4SRO overgrown by 2uc SIO and 2uc SZO. Our results strongly indicate that anomalies of the Hall resistance loops, which resemble a topological Hall effect contribution, may have other origins than non-trivial topological magnetic domains (e.g. isolated skyrmions or skyrmion lattices). Inhomogeneities of the local stoichiometry may result in variations of the local magnetization switching behavior and affect the anomalous Hall constant, and thus impact on the AHE resistivity loops measured while switching the magnetization of the samples between the saturated states.

\section{References}


\bibliographystyle{is-unsrt}
\bibliography{bibliography} 

\section{Acknowledgements} 
G.M., D.I., P.M. and L.M.E. gratefully acknowledge financial support by the German Science Foundation (DFG) through the Collaborative Research Center {\it Correlated Magnetism: From Frustration to Topology} (CRC1143) Project No. 247310070, the SPP2137 (Project No. EN 434/ 40-1), and projects No. EN 434/ 38-1 and No. MI 2004/ 3-1. L.M.E. also gratefully acknowledges financial support through the {\it Center of Excellence - Complexity and Topology in Quantum Matter} (ct.qmat) - EXC 2147. I.L-V. thanks the DFG for financial support (project LI3015/3-1, No. 335038432, and project LI3015/5-1, No. 403504808 within SPP 2137) and through the CRC1238.


\section{Competing financial interests}

The authors declare no competing financial interests.

\clearpage

\onecolumngrid
\setcounter{figure}{0}

\section*{Supplementary Online Material}


\renewcommand\thesubsection{S\arabic{subsection}}
\renewcommand{\thefigure}{S.\arabic{figure}}

\subsection{Pulsed Laser Deposition}
\label{PLD}

\begin{figure}[!ht]
	\includegraphics[width=\textwidth]{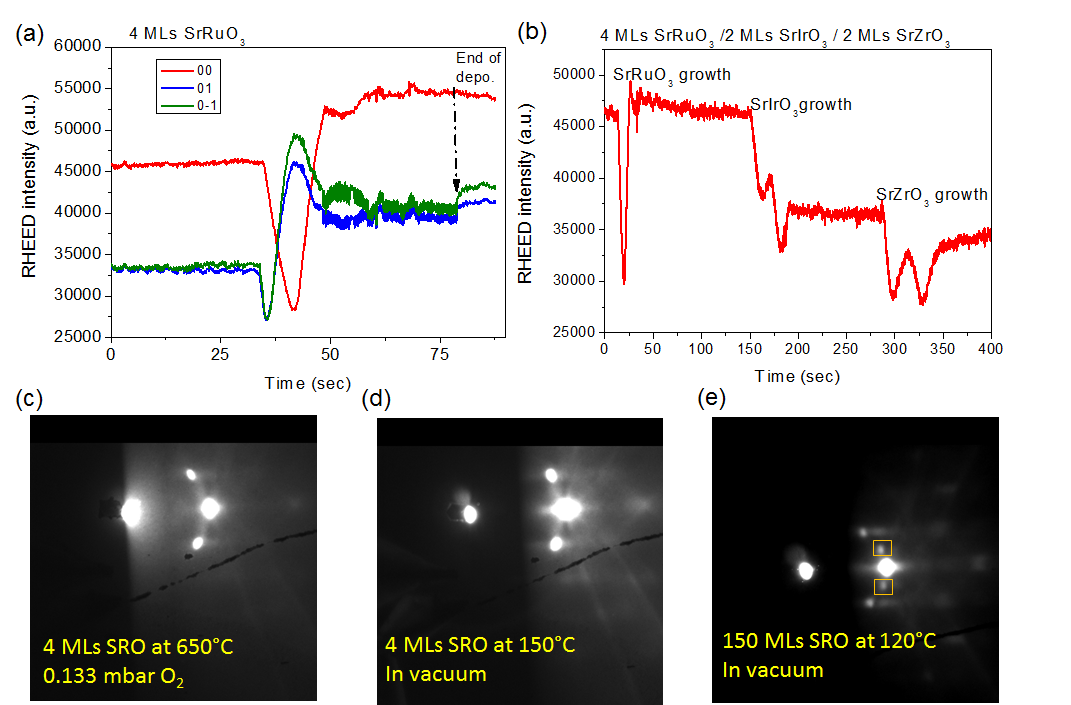}
	\caption{\label{PLDImage} RHEED intensity acquired during the growth of the (a) 4SRO and (b) of 4RIZ films. In (c) the RHEED pattern of the SRO film was taken under the growth conditions and in (d) the RHEED pattern of the same 4 uc SRO layer was acquired after the sample was cooled to $150^\circ\text{C}$ in 200 mbar O$_2$ and the chamber was evacuated. For comparison, in (e) the RHEED pattern of a 150 uc thick SRO layer is displayed, with the yellow boxes marking the extra reflections coming from the orthorhombic structure of the thick layer.}
\end{figure}

We monitored the PLD layer growth by reflective high energy electron diffraction (RHEED), in order to observe the growth mode and also to calibrate the layer thickness of the individual layers of the samples. Fig.~\ref{PLDImage} summarizes the RHHED investigations of both the 4SRO [Fig.~\ref{PLDImage}(a)] and 4RIZ [Fig.~\ref{PLDImage}(b)] sample, where the intensity of the RHEED specular spot (00) is plotted in red. These intensity behaviors indicate that the SRO layer in both samples was growing in a step-flow manner up to the 4uc nominal thickness, while the SIO and SZO layers for the 4RIZ sample grew in layer-by-layer mode to the desired 2uc thickness [two oscillations are monitored during SIO and SZO growth; see Fig.~\ref{PLDImage}(b)].

The RHEED patterns taken after the 4uc SRO layer growth at $650^\circ\text{C}$ and $150^\circ\text{C}$ indicated a very smooth growth pseudomorphic to the STO(100) substrate, also showing no orthorhombic reflections. In comparison, the RHEED pattern of a thick SRO film (150 MLs, i.e. about 60 nm thick) grown under the same conditions, shows orthorhombic reflections when cooled down to $120^\circ\text{C}$ [see Fig.~\ref{PLDImage}(c)].

\subsection{MOKE investigations}
\label{MOKE}

\begin{figure}[!htb]
	\includegraphics[width=0.5\textwidth]{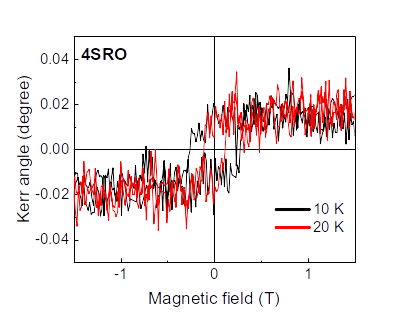}
	\caption{\label{MOKESRO} Polar magneto-optical Kerr effect measurements of the 4uc SRO film, performed with an incoherent light source of 600~nm wavelength: hysteresis loops of the Kerr rotation angle as a function of applied magnetic field at various temperatures.}
\end{figure}

The results of the magneto-optical Kerr effect investigations of the 4SRO film are summarized in Fig.~\ref{MOKESRO}. The squarish hysteresis loops are in good agreement with the sharp magnetization reversal observed in the microscopic MFM studies. Above 20~K, the background contribution is much larger than the signal originating from the bare 4 uc SRO film disabling a proper extraction of the sample contribution. 

\begin{figure}[!htb]
	\includegraphics[width=\textwidth]{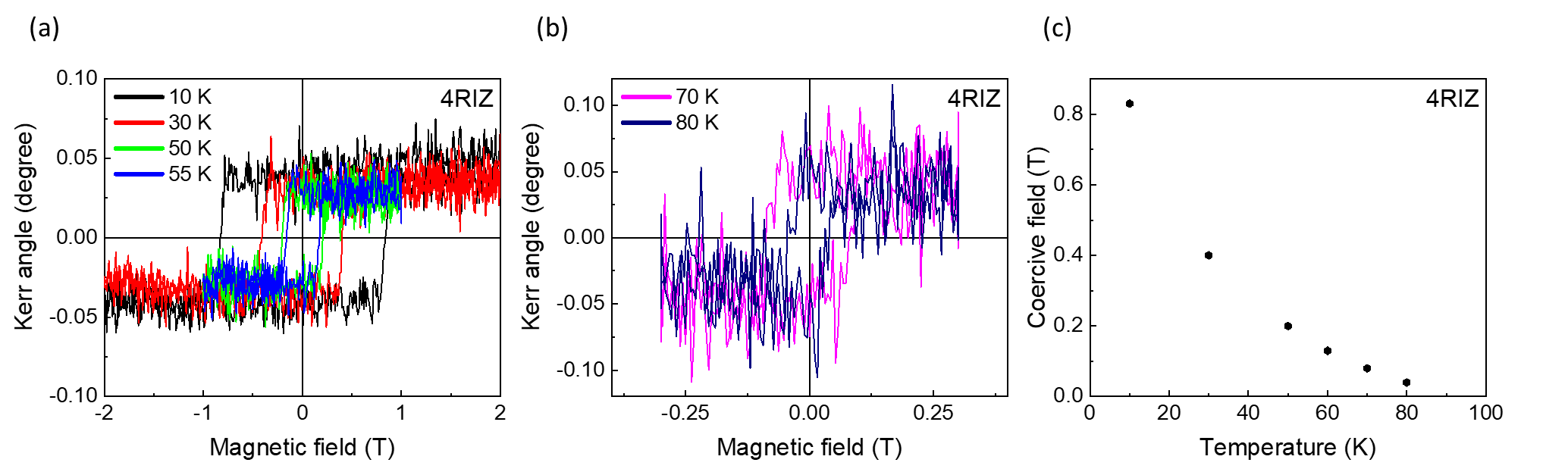}
	\caption{\label{MOKERIZ} Polar magneto-optical Kerr effect measurements of 4RIZ film, performed with an incoherent light source of 480~nm wavelength: (a)-(b) hysteresis loops of the Kerr rotation angle as a function of applied field at various temperatures; (c) Temperature dependence of the coercive field.}
\end{figure}

In Fig.~\ref{MOKERIZ} we summarize the magneto-optical Kerr effect studies of 4RIZ film. In accordance with the MFM investigations, the Kerr rotation angle loops show the switching behavior of a conventional ferromagnet for the magnetic field applied along the magnetic easy axis for the whole ferromagnetic temperature range. As shown in Fig.~\ref{MOKERIZ}(c), the coercive field strength exhibits a fast decrease with increasing temperature. The finite coercive field strength at 80 K is related to the enhancement of the ferromagnetic transition temperature compared with the bare SRO thin film.

\subsection{MFM measurement details on the 4SRO and 4RIZ thin-film samples}
\label{MFM}

\subsubsection{4SRO ultra-thin film at 10 K}
\begin{figure}[!htb]
	\centering
	\includegraphics[width=0.99\textwidth]{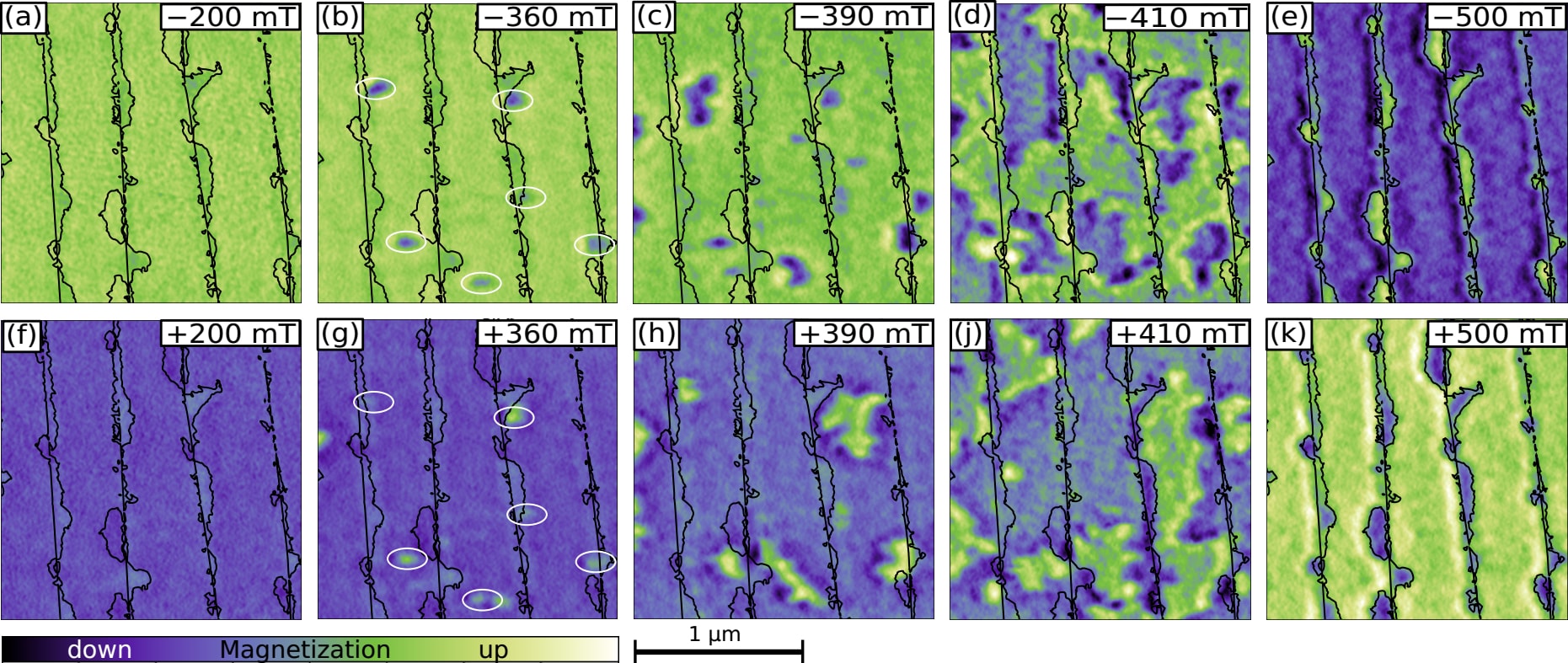}
	\caption{\label{figMagBare10Ka} MFM measurements of bare 4SRO ultra-thin films at 10~K, showing magnetic domain formation during forward magnetic field sweep (a)-(e) and reversed magnetic field sweep (f)-(k). Black lines mark step edges extracted from topography scan. Domain nucleation sites are marked in white in (b) and (g).}
\end{figure}

The series of MFM measurements of bare 4uc SRO ultra-thin film were performed down to the lowest temperature that is possible with our SFM setup, which is approximately 10~K. A series of relevant 10-K-MFM images as acquired while sweeping the external magnetic field from +2~T to -2~T, are displayed in Fig.~\ref{figMagBare10Ka}. At +2~T, the 4SRO sample is magnetically saturated. When reaching the zero field, no domain nucleation with opposite magnetization is observed, while the full SRO layer remains fully magnetized.

Nevertheless, a small but non-zero magnetic contrast is always visible at 0~T whenever comparing areas of a 4 and 3 uc height. Due to the different magnetic susceptibilities of these layers a clear, net magnetization for the 3 uc structure arises that in turn reveals the documented MFM contrast. 


At the negative field value of -60~mT, the tip changes its magnetization direction and aligns with the applied field. 

Starting at a field of -360~mT domains nucleate, with tiny round domains of less than 100~nm appearing, visible as dark round spots in Fig.~\ref{figMagBare10Ka}(b). For a field of -390~mT [Fig.~\ref{figMagBare10Ka}(c)] more domains appear and part of those already nucleated expand in size or coalesce with neighbors [Fig.~\ref{figMagBare10Ka}(d)]. In a field of -500~mT [Fig.~\ref{figMagBare10Ka}(e)] almost the entire area switched its magnetization, except a few domains which switch only in fields higher than -1~T. These domains correspond to the topographic features observed as islands formed on top of the terraces in Fig.~\ref{figTopo}(a), i.e. to areas of 5~uc thick SRO film.

Measurements performed on reversing the field between -2~T and positive fields exceeding the coercive field allowed us to observe memory effects in the nucleation of the domains. Comparing the  MFM images taken in a field of 360~mT [Fig.~\ref{figMagBare10Ka}(g)] and the MFM image scanned in -360~mT [Fig.~\ref{figMagBare10Ka}(b)], we observed that most of the nuclei appeared at the same sites. As not only these domains appear in the same positions, but these positions lie predominantly at step edges of either the SRO or STO surfaces, it has to be assumed that there is a strong preference for the nucleation to start at the sites of lattice defects.

From the switched magnetization versus unswitched magnetization area a coercive field value of about 400~mT was determined for 4~uc, see Fig.~\ref{figMagBare10Ka}(c). The thickness of 5~uc in SRO layer lead to sensitively higher coercive field above 1~T, that is close to the value in SIO/SRO/SZO trilayered heterostructure. The magnetization almost reached saturation for 500~mT [Fig.~\ref{figMagBare10Ka}(n)], only few domains with size smaller than 100~nm were unswitched, besides those domains related to the island in the topography, which switch in high fields.

\subsubsection{4SRO ultra-thin film at 55~K}
\begin{figure}[!htb]
	\includegraphics[width=0.5\textwidth]{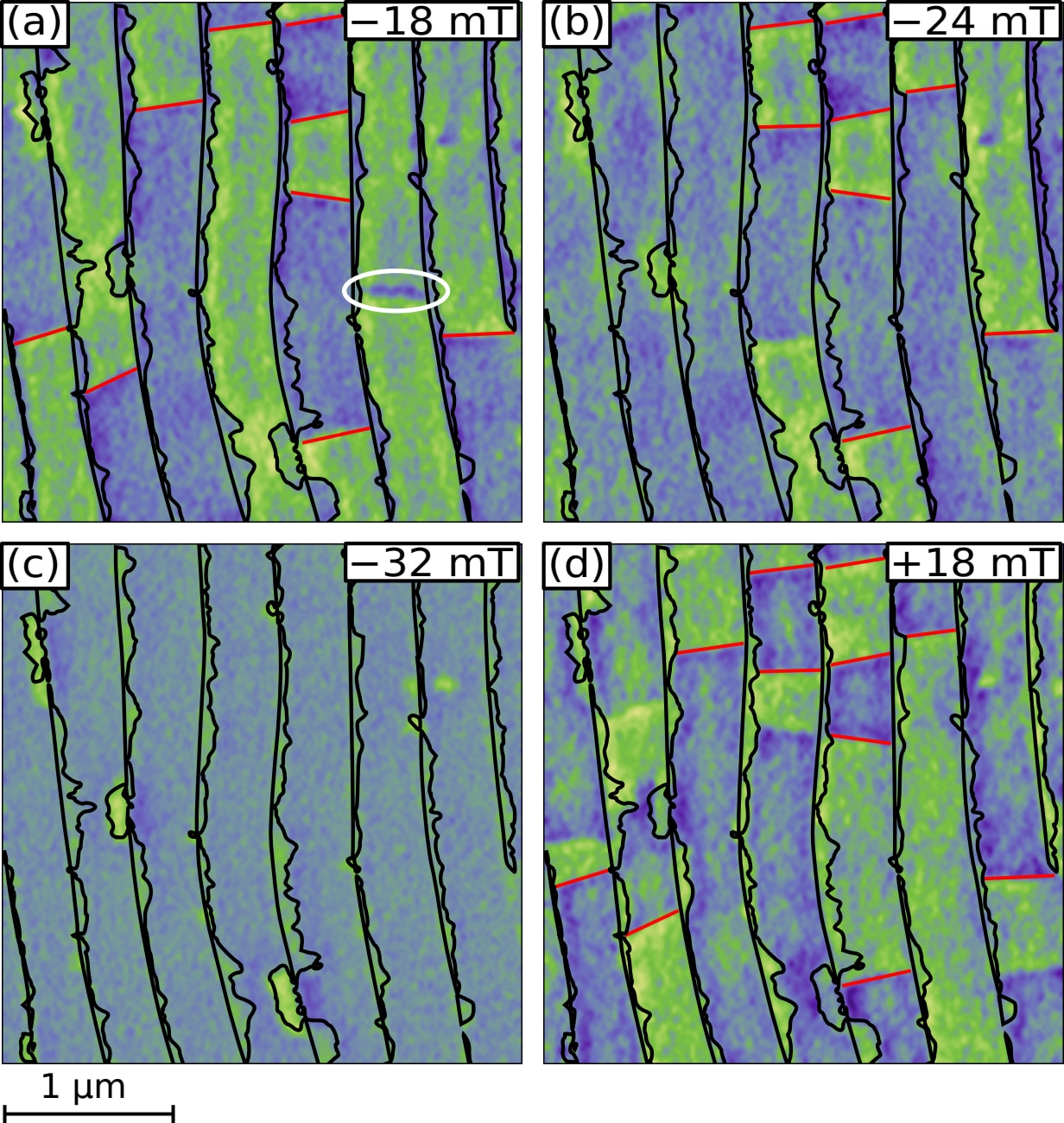}
	\caption{\label{figMagBare55Ka} Magnetic images of bare 4SRO ultra-thin film at 55~K. (a)-(c) MFM images of a sweep from positive to negative field. (d) Image from reverse sweep. The scan size is 3~$\mu$m x 3~$\mu$m. Red lines mark domain edge locations shared between pictures, black contours the topography steps.}
\end{figure}

Imaging at 55~K revealed similarities in the observed phenomenons of differing coercive fields for areas of differing film thickness and a strong pinning to specific sites. Representative images from the MFM field sweeps are shown in Fig.~\ref{figMagBare55Ka}.

The sample was magnetized in a high perpendicular magnetic field of 2~T and slowly switched to the low magnetic field of -2~T, and back. The coercive field is drastically reduced with the temperature, with the first domain of opposite magnetization is observed at applied field of -12~mT. Unlike for low temperatures, the shape and size of the nucleated domains (about 400~nm) match the terrace width of the substrate. All domains reach across the entire width of its step, even the line-shaped domain visible in Fig.~\ref{figMagBare55Ka}(a), marked with a white ellipse, strongly indicating that the nucleation of domains begins by a very rapid elongation of the initial nucleus perpendicular to the step edge.

While no memory effect could be observed for the initial nucleation due to the speed of domain growth, comparing the locations of domain walls across several images reveals a number of domain wall pinning sites, at which the growth stops for a time (Compare the red lines in Fig.~\ref{figMagBare55Ka}(d) to those in (a) and (b)).

For the areas with 5~uc of SRO, the coercive field remains higher than for the main terraces, although it is similarly reduced by the higher temperature, to a value of approximately 250~mT.

For a field value between -18~mT and -20~mT, the fraction of domains of magnetization aligned with the applied field direction reaches 50\%. Thus the value of the coercive field is a bit lower than 20~mT at 55~K. For a field value of -30~mT almost the entire scanned area switched its magnetization in the direction of the applied field. 

Two important conclusions can be drawn for the behavior of field-driven magnetic domains in bare SRO ultra-thin film at 55~K. Firstly, formation of the domains at this temperature is strongly influenced by the step-and-terrace morphology of the substrate. The size of nucleated domains matches the terrace width, and once formed these domains grow along the terraces and do not cross the surface steps. The 0.39~nm high surface steps appear to act as potential barrier for the domain wall propagation. 

Strong memory effects were exhibited as the domains nucleated at the same sites for both positive and negative field of about the same value. Secondly, at no stage of the magnetization reversal between saturation states did we observe the formation of either domains or skyrmions.
\\

\subsubsection{4SRO ultra-thin film at 80~K}
\begin{figure}[!htb]
	\includegraphics[width=0.5\textwidth]{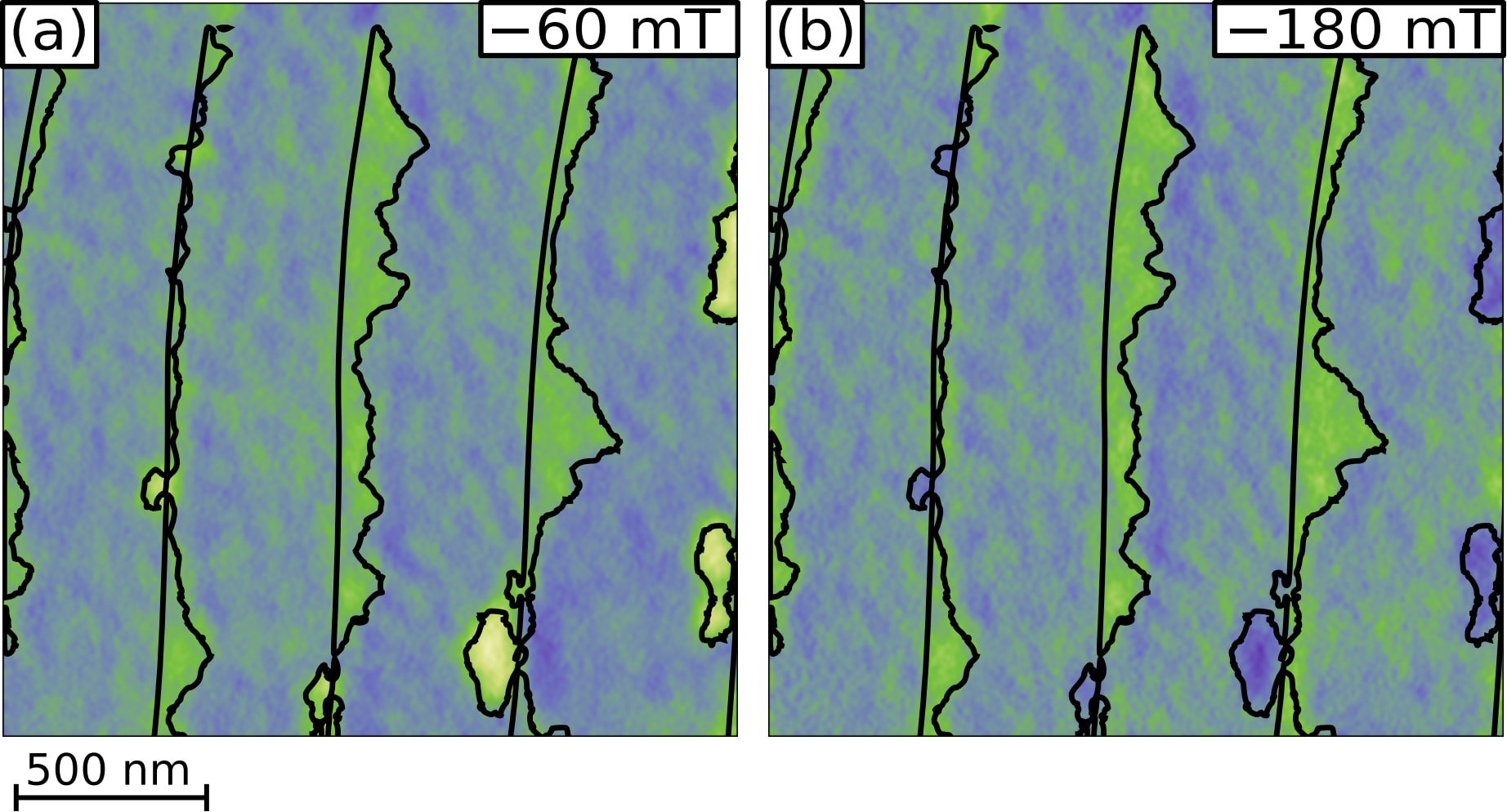}
	\caption{\label{Bare80Ka} MFM images of bare SRO ultra-thin film at 80 K. Only areas corresponding to an edge overgrowth show switching behaviour.}
\end{figure}

MFM investigations of bare SRO ultra-thin film at 80~K are presented in Fig.~\ref{Bare80Ka}. It shows that the 4~uc thick SRO terraces remain in its paramagnetic phase, while the 5~uc thick areas still display ferromagnetic behavior. It is the first important difference to the SIO/SRO/SZO trilayer. 

\subsubsection{4RIZ heterostructure at 10~K}
In contrast to the growth of the domains for the bare film, where the magnetic states of neighboring steps could be regarded as independent, the growth in the trilayer 4RIZ samples progresses both along and perpendicular to the steps. Similarly to the bare sample, the speed of the growth appears strongly dependent on local features, with swift growth to a fixed point followed by uncorrelated pauses of stagnation. Similarly, not every spot appears equally suitable for the cross-step growth, as the straight lines of step edges can still be observed as a strongly preferred domain edge.

\begin{figure}[!htb]
	\centering
	\includegraphics[width=0.5\textwidth]{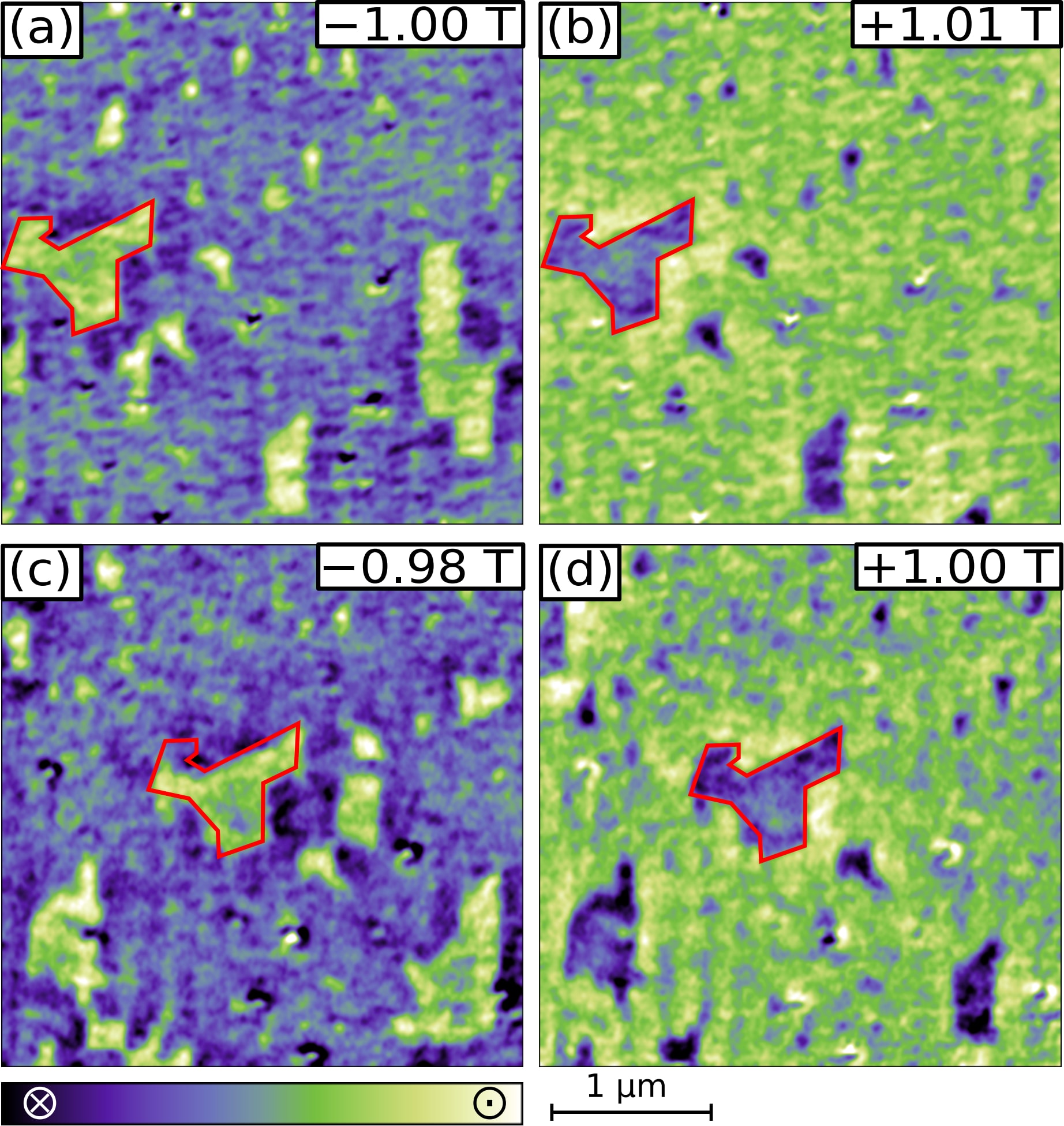}
	\caption{\label{Coated10K} MFM images of domains formed in the 4RIZ trilayer heterostracture at 10 K, with a recurring structure seen in several sweeps marked in red.}
\end{figure}

One example of pinning sites can be seen in Fig.~\ref{Coated10K}, which shows slices taken from successive sweeps of the same location on the crystal. In both directions and despite the long time (3 days) between those sweeps, a hammerhead shaped domain is visible, remaining stable against the magnetization reversal longer than the surrounding material. This structure spans several step edges in width and has several parallel boundary lines at an angle to the grid given by the vicinal surface, indicating a stronger influence of a different crystallographic direction on the domain wall stability at these spots.

\subsubsection{4RIZ heterostructure at 80 K}

\begin{figure}[!htb]
	\centering
	\includegraphics[width=0.5\textwidth]{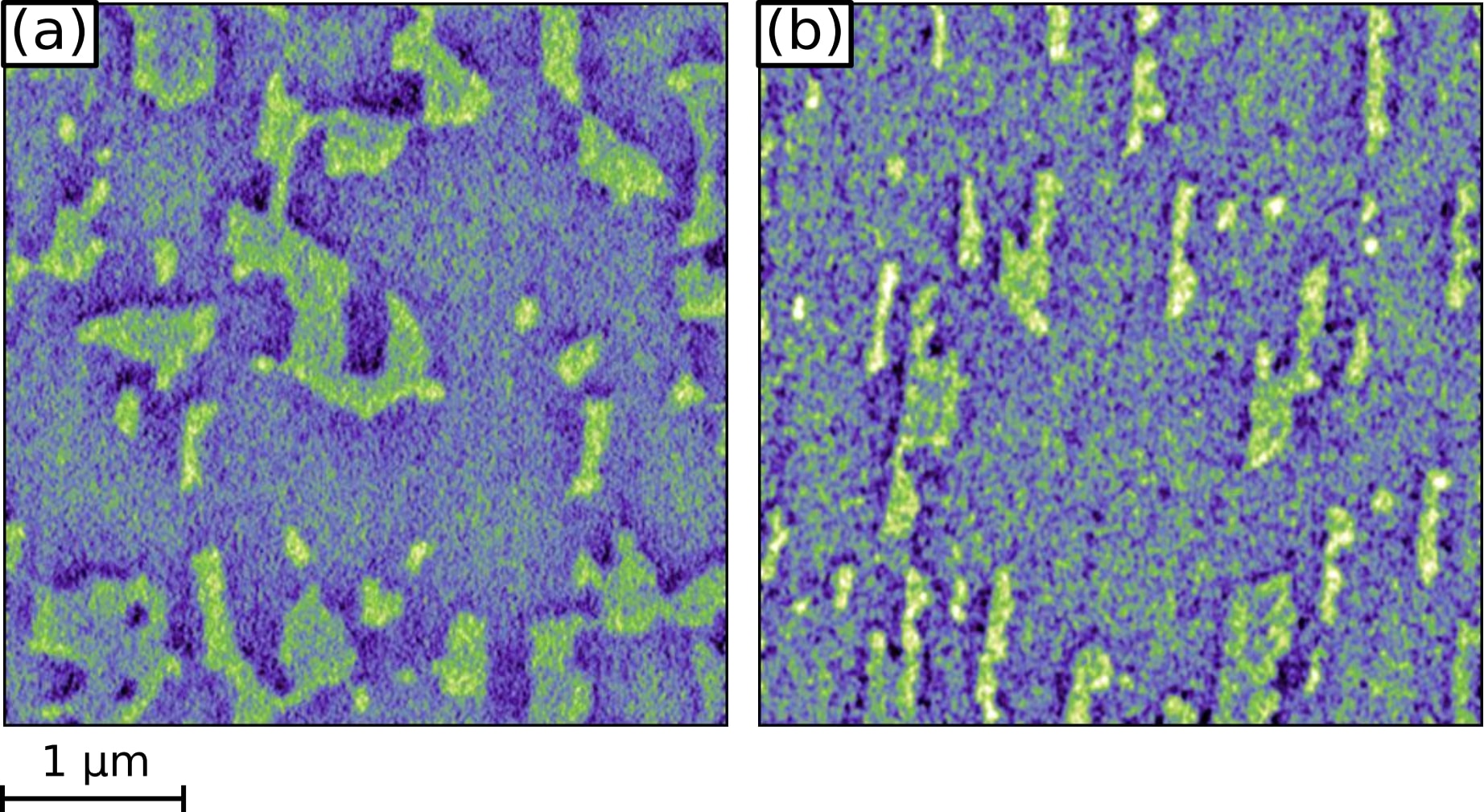}
	\caption{\label{CoatedFC} MFM images of the 4RIZ trilayer heterostructure at 80~K.  MFM phase images displaying: (a) magnetic domains formed upon zero-field cooling, (b) magnetic domains formed after prior saturation in -2~T, removal of the field and application of 40~mT magnetic field.}
\end{figure}

The magnetic domains in 4RIZ heterostructure formed upon zero-field cooled to 80~K, shown in the MFM image in Fig.~\ref{CoatedFC}(a). The domains have random shapes and sizes ranging from about 100~nm to a few $\mu$m. However, the domain patterns show a clear influence of the terrace ledges when they form after applying a large perpendicular magnetic field of 2~T, then removing the saturation field and applying a field of opposite polarity (values of 40-50~mT, which exceeded the coercive field at 80 K). The domains formed follow the direction of the terraces, being elongated parallel to the ledges and extended in widths primarily across a single terrace (200-450~nm width), as is shown in Fig.~\ref{CoatedFC}(b).

\subsection{Local-scale MFM hysteresis loops and Otsu's method}
\label{Hysteresis}

\begin{figure}[!ht]
	\centering
	\includegraphics[width=\textwidth]{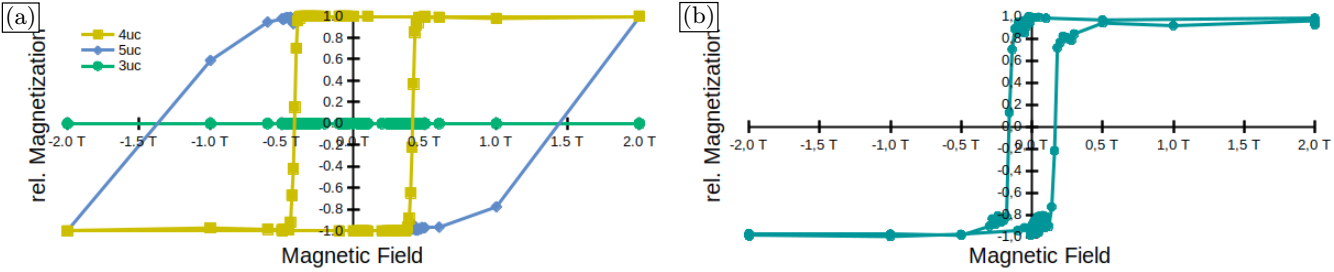}
	\caption{\label{Hystereses} Hystereses obtained from MFM data, for the bare sample at 10~K (a) and the trilayer at 55~K (b).}
\end{figure}


To extract magnetization value and therefore hysteresis from the MFM images, an thresholding approach is used: A threshold value is chosen for the image such that the upwards magnetized areas are above this value in signal, and the downwards magnetized areas below. Then, the total magnetization, as a ratio compared to the saturation magnetization, can be simply obtained as a ratio of the thus selected areas.

The thresholding is automated using the optimization measure invented by Noboyuki Otsu~\cite{Otsu79}, splitting the image such that the sum of weighted variances of the resulting parts is minimal. This algorithm is known to produce good results for images containing two classes of brightness, even if the width of the classes is so large as to cause significant overlap in the histogram.

For the initial saturated state and the onset of nucleation, the assumptions for the validity of Otsu's method are not fulfilled, resulting in the algorithm giving a nonsensical value for the magnetization based on its attempt to separate the measurement noise on top of the constant saturation value. To correct for these cases, additional fixed thresholds were chosen above and below the average value of the image, with their value corresponding to the greatest distance of Otsu's threshold value to the image average. These thresholds are sensitive to the initial nucleation and final domains respectively, allowing for the total hysteresis to be stitched together from the three curves by merging them at their tangent points.

While the method can only be applied for ferromagnetic transitions and is insensitive to paramagnetic responses, it at the very least gives the correct value for the coercive field. In addition, as it is based on the MFM images, it can be used to spatially resolve this field, by determining when any individual pixel switches magnetization, as determined by its progress over the relevant threshold.

\end{document}